\documentclass[sigconf]{acmart}

\AtBeginDocument{%
  }

\copyrightyear{2026}
\acmYear{2026}
\setcopyright{cc}
\setcctype{by}
\acmConference[CHI '26]{Proceedings of the 2026 CHI Conference on Human Factors in Computing Systems}{April 13--17, 2026}{Barcelona, Spain}
\acmBooktitle{Proceedings of the 2026 CHI Conference on Human Factors in Computing Systems (CHI '26), April 13--17, 2026, Barcelona, Spain}
\acmPrice{}
\acmDOI{10.1145/3772318.3790533}
\acmISBN{979-8-4007-2278-3/2026/04}

\begin{document}

\title{Identifying, Explaining, and Correcting Ableist Language with AI}

\author{Kynnedy Simone Smith}
\orcid{0009-0008-0857-1277}
\affiliation{
  \department{Human-Computer Interaction Institute}
  \institution{Carnegie Mellon University}
  \city{Pittsburgh}
  \state{Pennsylvania}
  \country{USA}
}
\email{kynnedys@andrew.cmu.edu}

\author{Lydia B Chilton}
\orcid{0000-0002-1737-1276}
\affiliation{
  \department{Computer Science Department}
  \institution{Columbia University}
  \city{New York}
  \state{New York}
  \country{USA}
}
\email{chilton@cs.columbia.edu}

\author{Danielle Bragg}
\orcid{0000-0002-7846-3481}
\affiliation{
  \institution{Microsoft Research}
  \city{Cambridge}
  \state{Massachusetts}
  \country{USA}
}
\email{dbragg@post.harvard.edu}


\begin{abstract}
Ableist language perpetuates harmful stereotypes and exclusion, yet its nuanced nature makes it difficult to recognize and address. Artificial intelligence could serve as a powerful ally in the fight against ableist language, offering tools that detect and suggest alternatives to biased terms. This two-part study investigates the potential of large language models (LLMs), specifically ChatGPT, to rectify ableist language and educate users about inclusive communication. We compared GPT-4o generations with crowdsourced annotations from trained disability community members, then invited disabled participants to evaluate both. Participants reported equal agreement with human and AI annotations but significantly preferred the AI, citing its narrative consistency and accessible style. At the same time, they valued the emotional depth and cultural grounding of human annotations. These findings highlight the promise and limits of LLMs in handling culturally sensitive content. Our contributions include a dataset of nuanced ableism annotations and design considerations for inclusive writing tools.
\end{abstract}

\ccsdesc[500]{Human-centered computing~Empirical studies in HCI}
\ccsdesc[300]{Human-centered computing~Participatory design}
\ccsdesc[100]{Human-centered computing~Natural language interfaces}
\ccsdesc[500]{Social and professional topics~People with disabilities}
\ccsdesc[500]{Computing methodologies~Discourse, dialogue and pragmatics}





\maketitle

\section{Introduction}
Bias in language is a powerful force that shapes perception and reinforces inequality. While overt forms of discrimination such as racial slurs or sexist insults are often condemned and filtered out by both social norms and algorithmic safeguards, more subtle forms of bias continue to persist and cause harm \cite{blodgettLanguageTechnologyPower2020,schmidtSurveyHateSpeech2017,fortunaSurveyAutomaticDetection2019, tylerDisAbilityStudies2015}. One such form is ableism, defined as the discrimination of and social prejudice against people with disabilities, grounded in the belief that non-disabled ways of being are superior \cite{nationalcenterondisabilityandjournalismDisabilityLanguageStyle2021,americanpsychologicalassociationInclusiveLanguageGuide20233,americanpsychologicalassociationwashingtondistrictofcolumbiaPublicationManualAmerican2020}. Like racism and sexism, ableism devalues entire groups, but it often manifests through casual, everyday expressions that go unnoticed by those outside the disability community \cite{bottema-beutelAvoidingAbleistLanguage2021,garland-thomsonFeministDisabilityStudies2005,hassanUnpackingInterdependentSystems2021,narayananvenkitAutomatedAbleismExploration2023,steenGlossaryAbleistWords2024, tylerDisAbilityStudies2015}.

This paper focuses on nuanced ableism, or subtle language that perpetuates negative assumptions, stereotypes, or limitations about disability without using explicit slurs or insults. Examples include phrases like “suffers from [insert disability]” or “wheelchair-bound,” which may seem neutral to some but frame disability as inherently pitiable, burdensome, or restrictive \cite{campbellContoursAbleismProduction2009, tylerDisAbilityStudies2015,nationalcenterondisabilityandjournalismDisabilityLanguageStyle2021,americanpsychologicalassociationInclusiveLanguageGuide20233,americanpsychologicalassociationwashingtondistrictofcolumbiaPublicationManualAmerican2020}. Despite growing awareness around inclusive language, these subtle forms of ableism remain widespread in daily communication, media coverage, and even academic and professional writing \cite{kaferFeministQueerCrip2013,americanpsychologicalassociationwashingtondistrictofcolumbiaPublicationManualAmerican2020}.

While all forms of bias deserve scrutiny, ableism presents unique challenges for both detection and intervention. Compared to racism or sexism, ableist language is less publicly recognized, more socially normalized, and often deeply entangled with medicalized or pity-based assumptions about disability \cite{campbellContoursAbleismProduction2009,tylerDisAbilityStudies2015}. This makes ableist expressions less likely to be flagged by general hate-speech tools and more likely to be overlooked even by well-intentioned communicators.

Human expertise remains central to identifying and interpreting ableist language. However, exclusive reliance on experts or disabled advocates does not scale to the volume and pace of everyday written communication. Even when compensated, expecting disability community members to continuously correct others’ language reinforces an inequitable division of labor \cite{chordiaSocialJusticeHCI2024,pierreGettingOurselvesTogether2021}. Moreover, prior research in related bias domains shows that automated tools can function as supportive first-pass systems that reduce both cognitive load and emotional labor for marginalized communities \cite{blodgettLanguageTechnologyPower2020,fortunaSurveyAutomaticDetection2019}.

We therefore explore AI-based assistance not as a replacement for disabled expertise, but as a scalable, widely accessible complement. Culturally competent AI has the potential to support everyday writing tools, journalism and document style checking, educational training environments, and professional communication systems where rapid, context-sensitive feedback can prevent harm before publication. Lessons from other domains of harmful language detection — such as racist or sexist microaggressions — demonstrate that tailored datasets and evaluation frameworks are necessary to address the specific cultural contexts of each form of bias \cite{narayananvenkitAutomatedAbleismExploration2023}. Ableism requires similar dedicated attention.

This paper explores the current capabilities of large language models, specifically GPT-4o, in identifying, correcting, and explaining ableist language. We investigate the following research questions:

\begin{itemize}
    \item RQ1: Do humans prefer AI annotators or human annotators for identifying, explaining, and correcting ableist language?
    \item RQ2: What qualities of AI and human ableism annotations make them agreeable or disagreeable to the participants?
    \item RQ3: How can AI annotators improve at identifying, explaining, and/or correcting ableist language?
\end{itemize}
We investigate these research questions in a two-part study. In part 1, we investigate how members of the disability community identify, explain, and correct ableist language in identity-specific short studies. We use the data from this study to create a human crowdsourced annotation that is used in part 2, where we survey 100+ other members of the disability community to compare the human crowdsourced annotation with an AI annotator. Participant feedback provides insights on the strengths, weaknesses, and areas of improvement for both annotations. From these studies, we contribute:

\begin{enumerate}
    \item \textbf{\textbf{A dataset of nuanced ableism annotations from the disability community.}}
    \item \textbf{\textbf{An exploration of the tradeoffs between AI and human crowdsourced annotations of nuanced ableism.}}
    \item \textbf{\textbf{A list of considerations for developers of LLMs and writing tools that deal with culturally sensitive language.}}
\end{enumerate}
By advancing our understanding of AI’s role in promoting inclusive language, this work aims to foster more effective and widespread use of culturally competent communication tools that allow people to participate using inclusive language in their daily lives.

\section{Related Work}

\subsection{Disability Justice and Foundations of Ableist Language}

Disability studies and disability justice frameworks emphasize that ableism is a historically rooted system of oppression that privileges normative culture and marginalizes people with disabilities \cite{campbellContoursAbleismProduction2009,garland-thomsonFeministDisabilityStudies2005,tylerDisAbilityStudies2015}. Disability Justice principles emphasize intersectionality awareness, collective access, and the centering of those most affected in decision-making \cite{berneTenPrinciplesDisability2018}. These frameworks highlight that language interventions are never neutral: they can either reinforce exclusion or support disabled autonomy.

Research in this space has also documented the prevalence of subtle, implicit ableism—microaggressions that pathologize, infantilize, or erase disability even when overt slurs are absent \cite{keller_microaggressive_2010,bottema-beutelAvoidingAbleistLanguage2021}. Language guidance from the Center for Disability Journalism and the APA reinforces that stigma often manifests through framing and metaphor, not merely harmful words \cite{americanpsychologicalassociationInclusiveLanguageGuide20233,nationalcenterondisabilityandjournalismDisabilityLanguageStyle2021}.

Yet most language frameworks remain prescriptive checklists that cannot adapt to reclaiming practices, cultural differences, or narrative context \cite{kaferFeministQueerCrip2013,gadirajuWouldntSayOffensive2023}. Our work contributes by examining not only \textit{what} language is considered ableist, but \textit{how} disabled participants explain and propose alternative representations—data rarely formalized in existing approaches.

\subsection{Ableism in AI Systems and Human–AI Disagreement}

Scholars have warned that AI systems can reinforce ableist norms through biased datasets and incomplete representations of disability \cite{whittaker_disability_2019,odeaOutBoundsFactors2015}. Recent HCI and NLP work shows that models attribute unwarranted sentiment to disability terms \cite{narayananvenkitAutomatedAbleismExploration2023}, misinterpret community language \cite{heungVulnerableVictimizedObjectified2024}, and produce dehumanizing stereotypes \cite{mackTheyOnlyCare2024}. These harms often arise not from explicit hate but from implicit assumptions embedded in everyday text \cite{keller_microaggressive_2010,heungVulnerableVictimizedObjectified2024}.

A growing line of research directly evaluates how AI explains or critiques disability-related harm. Phutane et al. show that AI explanations of ableism are “cold and condescending,” diverging from disabled participants in moral framing and detail \cite{phutaneColdCalculatedCondescending2025}. Our focus extends this work by examining not only differences in explanation, but also comparison of identification and correction tasks—providing a broader view of annotation alignment.

Similarly, prior guidelines call for participatory disability-centered pipelines \cite{liangEmbracingFourTensions2021,pierreGettingOurselvesTogether2021}, yet their evaluations rarely examine downstream user experience. We contribute empirical evidence of how disabled readers perceive AI feedback—highlighting where AI informs, harms, or overreaches.

\subsection{Automated Detection and Correction of Biased Language}

A substantial body of work addresses toxicity, hate speech, and bias classification \cite{schmidtSurveyHateSpeech2017,fortunaSurveyAutomaticDetection2019,gordonDisagreementDeconvolutionBringing2021}. However, systems optimized for safety often penalize identity terms or reclaimed language, leading to over-moderation of marginalized users \cite{heungVulnerableVictimizedObjectified2024}. Researchers therefore propose explanatory systems that articulate social power dynamics and guide inclusive alternatives \cite{sapSocialBiasFrames2020,hassanUnpackingInterdependentSystems2021}, and motivate AI that supports learning rather than punishment.

Despite progress in racism and sexism detection, scholarship on automated ableism annotation remains limited and is often related to toxicity or bias scores \cite{glazkoIdentifyingImprovingDisability2024,liDecodingAbleismLarge2024a,phutaneColdCalculatedCondescending2025,narayananvenkitAutomatedAbleismExploration2023,fortunaSurveyAutomaticDetection2019}. And while rewriting systems have shown promise in reducing harmful phrasing \cite{jangConsistencyAnalysisChatGPT2023}, little is known about how users interpret AI-proposed revisions—especially when disability representation is deeply tied to narrative and identity. Prior ableism-focused work has primarily examined classification accuracy, not perceived clarity, authority, or emotional impact.

Our work advances this space by evaluating not only whether AI can detect and revise subtle ableism, but how those revisions are interpreted by the disabled communities they aim to serve. By comparing human and AI-generated annotations side-by-side (notably without consideration of a toxicity or ableism score), we move beyond automated performance metrics toward understanding user trust, clarity, and alignment with disability justice values.

\section{Study 1 - Collecting Annotations from the Disability Community}

The goal of Study 1 was to examine how disability community members interpret and annotate ableist content, and to establish a set of human baseline annotations for later comparison with AI-generated annotations. We sought to surface both points of consensus and contestation, illuminating the challenges of identifying nuanced ableism.

\subsection{Methods}

We used GPT-4o to generate one-paragraph fictional stories about individuals with the following seven disabilities: vision impairment, hearing impairment, mental health conditions, intellectual or learning disabilities, neurological disabilities, autism, and reduced mobility. Prior research has shown that large language models (LLMs) produce subtle and overt ableist content when prompted with disability-related topics \cite{mackTheyOnlyCare2024,gadirajuWouldntSayOffensive2023}, and generating the texts ourselves enabled consistency across disability groups. The disability categories match the long-term disability screening options provided by our survey platform, Prolific, which enabled us to recruit experts to evaluate each story type.

The prompt used to generate each story was: \textit{“Write a 5 to 8 sentence short story about a [insert disability here] doing their job or completing a task.”} This prompt comes from Gadiraju et. al's paper, which showed that ChatGPT produces ableist stories when prompted to write stories about people with disabilities doing things like completing tasks or doing a job \cite{gadirajuWouldntSayOffensive2023}. We required the stories to be within five to eight sentences long to control the length of the annotation survey. To ensure the generated stories included nuanced ableism, we piloted the texts with 3 individuals from each corresponding disability group. These participants were asked to verify that the texts satisfied a list of criteria formulated from anti-bias and ableism guides \cite{departmentofglobalcommunicationsDisabilityInclusiveCommunicationsGuidelines2022,americanpsychologicalassociationInclusiveLanguageGuide20233,americanpsychologicalassociationwashingtondistrictofcolumbiaPublicationManualAmerican2020,nationalcenterondisabilityandjournalismDisabilityLanguageStyle2021,steenGlossaryAbleistWords2024}. Please see the Appendix for the list of criteria. We generated stories using the same prompt until at least 2/3 pilot participants agreed that ableism following our guidelines was present. 

After generating ableist stories, we conducted a study to investigate how in-group experts annotated these stories for ableist concepts. We later utilize the data from this study to create 7 different human baseline annotations to compare against our AI annotations in study 2.

\subsubsection{\textbf{Implications of AI-Generated Stories}}


Our decision to use GPT-4o to generate narrative prompts was driven by both practical and methodological considerations. This study focuses on nuanced forms of ableism that emerge through storytelling—via tone, character framing, and implication—rather than through isolated phrases alone. While short-form text, such as the social media comments evaluated in Phutane et al. \cite{phutaneColdCalculatedCondescending2025}, can capture instances of implicit ableism, these contexts often make it difficult to offer meaningful corrections without altering the speaker’s underlying intent.

For example, the phrase from Phutane et al.'s dataset, “It’s so inspiring that you can make your own coffee”\cite{phutaneColdCalculatedCondescending2025}, is implicitly ableist when directed toward a person with a disability, and both humans and AI can readily explain why. 
For example, the phrase from the dataset by Phutane et al., “It’s so inspiring that you can make your own coffee”\cite{phutaneColdCalculatedCondescending2025}, is implicitly ableist when directed toward a person with a disability, and both humans and AI can readily explain why. However, correcting such a comment without reframing or negating the speaker’s intended compliment is challenging; attempts to preserve the positive intent often reproduce the same underlying assumption that everyday activities are exceptional for disabled individuals.

In contrast, narrative contexts allow subtle ableism to unfold across a coherent arc and provide opportunities for corrections that are more likely to preserve the author’s broader intent. For instance, in our dataset, the line “As the aroma of freshly brewed coffee filled the room, Emily felt a sense of accomplishment” was labeled as ableist by 33\% of participants in Study 2, who cited excessive praise, infantilization, and lowered expectations regarding the quality of life of Emily, a blind character. Because this phrasing appears as a detail within a larger narrative, corrections can shift “a sense of accomplishment” to emotions such as calm, anticipation, or preparedness—maintaining the narrative’s structure while removing the ableist implication. This affordance makes narrative text particularly well-suited for studying not only the identification of subtle ableism, but also how it can be meaningfully corrected. The full story used in this example can be found in the Appendix.

Using AI to produce these narratives also ensured standardization across disability types. Every participant annotated a story with comparable complexity, length, and structure, which would be extremely difficult to source or commission in a manner that preserves equity across seven disability groups. Piloting with in-group participants further ensured that stories contained community-recognized ableism rather than artificially inserted slurs.

At the same time, this design choice necessarily shapes the interpretive context: the stories reflect ableist patterns that LLMs tend to generate. Although this approach was intentional—building on evidence that LLMs reproduce both subtle and overt ableism \cite{narayananvenkitAutomatedAbleismExploration2023,heungVulnerableVictimizedObjectified2024}—it limits generalizability. Additional work should test whether results hold for human-authored narratives that reflect broader diversity in writing styles, values, and cultural expression.

Overall, AI-generated stories offered an ethically and logistically feasible starting point for studying narrative ableism in a controlled way, while also motivating future research on more varied and community-produced texts. All of the AI-generated stories used in this study can be found in the supplemental information section and online (\href{https://doi.org/10.5281/zenodo.18386877}{link}) \cite{smith2026supplementary}.

\subsubsection{\textbf{Participants}}

\begin{table}[h]
\centering
\caption{Participant Demographics (N = 110)}
\label{tab:participants}
\begin{tabular}{lp{4.5cm}}
\toprule
\textbf{Category} & \textbf{Participant Data} \\
\midrule
Total Participants & 110 \\
\addlinespace[0.5em]
Age Distribution &
18--25: 15 \newline
26--35: 36 \newline
36--45: 27 \newline
46--55: 20 \newline
56--65: 8 \newline
66+: 4 \\
\addlinespace[0.5em]
Gender Distribution &
Female: 51\% (56) \newline
Male: 36\% (40) \newline
Non-binary/third gender: 10\% (11) \newline
Other: 1.8\% (2) \newline
Prefer not to say: 0.9\% (1)\\
\addlinespace[0.5em]
Disability Distribution &
Mental Health Condition: 74 \newline
Physical Disability / Reduced Mobility: 44 \newline
Autism: 36 \newline
Vision Impairment: 35 \newline
Neurological Disability: 35 \newline
Intellectual or Learning Disability: 27 \newline
Hearing Impairment: 26 \\
\addlinespace[0.5em]
Percentage who Identify \\with Multiple Disabilities & 75.45\% \\
\bottomrule
\end{tabular}
\end{table}

We used Prolific to recruit 110 participants who self-identified with the disability community reflected in their assigned survey. Each participant had completed at least a bachelor’s degree and had prior experience with diversity, equity, and inclusion (DEI) or disability-related training or coursework.

Participants completed one of seven survey tracks based on disability identification: Vision Impairment, Hearing Impairment, Mental Health Condition, Intellectual or Learning Disability, Neurological Disability, Autism, and Physical Disability / Reduced Mobility. Each of the seven surveys had 15-18 responses. The most common age range among participants was 26–35 years, and the majority identified as female. The most frequently reported disabilities were mental health conditions, physical disabilities or chronic illnesses, and autism. To account for many people's intersectional disabilities \cite{crenshawMappingMarginsIntersectionality1991,tinagoethalsWeavingIntersectionalityDisability2015,garland-thomsonFeministDisabilityStudies2005} and ensure a unique set of participants for each study, participants were only eligible to opt-in to one of our surveys.

Participants received \$4.50 for completing the 15-minute survey, with the average completion time around 18 minutes. Those who spent additional time ($\leq$ 5 extra minutes) received bonuses of up to \$4 for their effort. This was more common for users who use assistive technology to access our survey, like members from the blind community and the reduced mobility community. Our compensation aligned with Prolific’s recommended fair-pay rates and was designed to scale with effort (base rate + bonus for extended annotation).

\subsubsection{\textbf{Study Design}}

\begin{figure*}[h!]
    \centering
    \includegraphics[width=0.4\linewidth]{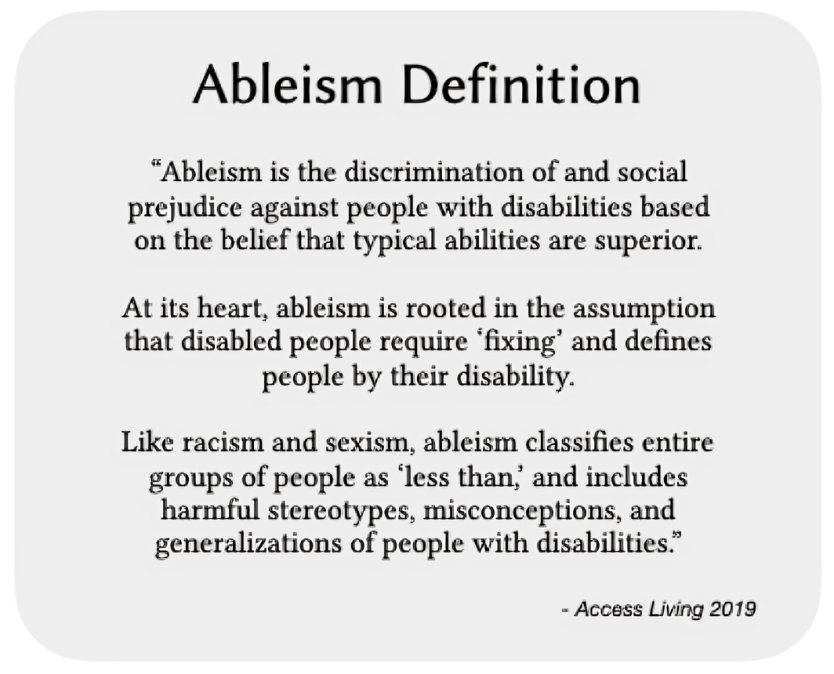}
    \caption{We use this definition of ableism in our user studies and throughout this paper. It was created by Ashley Eisenmenger, a disability inclusion training specialist at Access Living.}
    \label{fig:Ableism Definition}
    \Description{Figure 1. Fully described in the text.}
\end{figure*}

After providing consent and demographic information, participants ranked their agreement with the definitions of "ableism" created by Access Living\footnote{Access living is a leading disability advocacy organization based in Chicago. See their definition of Ableism here: https://www.accessliving.org/newsroom/blog/ableism-101/} and the definition of "nuanced ableism" we developed by synthesizing existing guidelines from the UN disability inclusive language guidelines, the APA Inclusive Language Guide, and the National Center on Disability and Journalism style guide into a framework for this study. These definitions (see Figure 1 and Appendix A) grounded participants before annotation. Participants had the option to provide rationale if they disagreed with the provided definitions, but all participants replied "somewhat agree" or "strongly agree" with the terms.

Next, participants completed a tutorial on our three-part annotation task: identification, explanation, and correction. The tutorial included examples of strong and weak annotations and utilized three attention checks. Participants needed to pass at least two of these checks to proceed with the survey.

After the tutorials, participants started the annotation section of the survey. First, they read the short fictional story related to their disability group. Next, they were presented with each sentence of the story and used a Likert scale to indicate how much they agreed with the statement, “The above sentence has ableist content”. If a participant selected strongly agree, somewhat agree, or neither agree nor disagree, they were prompted to annotate the sentence for ableism in three steps:
\begin{enumerate}
    \item Identification: “What words or phrases do you believe make this sentence ableist? Write only the exact words or phrases that appear in the sentence and separate phrases with commas."
    \item Explanation: “In less than 300 characters (2-3 sentences), give a brief explanation of why the words or phrases make the sentence ableist. Use an objective and educational tone like a spell checker would. Refrain from speaking from the "I" perspective and giving personal examples/anecdotes."
    \item Correction: "Rephrase the sentence using inclusive alternatives to the ableist phrase(s) you identified. Write the full corrected sentence and nothing else."
\end{enumerate}

We only collect written responses for a sentence if a participant believed it contained ableism. Adding written responses for sentences without ableism would have drastically increased cognitive load and average completion time. After completing the annotation task for each sentence of the story, participants were asked to provide an \textbf{overall analysis} of ableism in the story. If they were neutral or agreed with the statement, "Do you believe that concepts of the overall story contain explicit and/or nuanced ableist content?" they followed similar annotation steps:
\begin{enumerate}
    \item Identification: “What overall concepts in the story are ableist? If there is more than one ableist concept, separate it with commas."
    \item Explanation: “Explain why the concepts you identified are ableist:"
    \item Correction: "How could the story be reworked to avoid integrating the ableist concepts you identified?"
\end{enumerate}

This two-level annotation approach (\textbf{sentence-level and passage-level}) enabled us to capture both micro-linguistic and macro-narrative manifestations of ableism, which prior work has often treated separately.

\subsection{Results}

Findings from Study 1 highlight the subjectivity related to identifying ableist language, particularly when it is nuanced or context-dependent. While inter-annotator agreement was generally low, the majority of participants still recognized overarching ableist themes across the stories, validating the design of our stimuli. 

\subsubsection{\textbf{Annotation Collection}}

We collected a total of 276 ableism annotations across seven surveys administered to 110 participants. Of the 276 annotations collected, 210 were sentence-level annotations and 66 reflected passage-level thematic concerns. The volume of sentence annotations varied across disability categories represented in the stories. Table 2 shows the distribution of sentence-level annotations by disability category.

\begin{table*}[h]
\centering
\caption{Number and Percentage of Sentence- and Story-Level Annotations by Disability Community}
\label{tab:annotations}
\begin{tabular}{lcc}
\toprule
\textbf{Disability Community Survey} & \textbf{Sentence-Level Annotations} & \textbf{Story-Level Annotations} \\
\midrule
Intellectual/Learning & 48 (22.9\%) & 14 (21.2\%) \\
Reduced Mobility & 44 (21.0\%) & 12 (18.2\%) \\
Neurodiversity & 37 (17.6\%) & 10 (15.2\%) \\
Deaf & 25 (11.9\%) &  9 (13.6\%) \\
Autism & 22 (10.5\%) &  5 (7.6\%) \\
Blind & 18 (8.6\%)  &  8 (12.1\%) \\
Mental Health & 16 (7.6\%)  &  8 (12.1\%) \\
\midrule
\textbf{Total} & \textbf{210}& \textbf{66}\\
\bottomrule
\end{tabular}
\Description{Table 2. Fully described in the text.}
\end{table*}

Stories concerning intellectual disabilities, mobility, and neurodiversity received the highest number of sentence-level annotations, suggesting that these forms of ableism were more frequently recognized. A core contribution of this paper is an annotated dataset of all 276 Annotations collected in this study.

\subsubsection{\textbf{Inter-Annotator Agreement}}

We assessed inter-annotator agreement using Fleiss' Kappa (FK) and Krippendorff's Alpha (operationalized as Kendall's Tau due to ordinal data). Agreement scores varied by disability type, reflecting the subjective nature of interpreting subtle forms of ableism. Results are summarized in Table 3.

\begin{table*}[h]
\centering
\caption{Inter-annotator agreement across disability categories using Fleiss’ Kappa (FK) and Krippendorff/Kendall’s Tau.}
\label{tab:agreement}
\begin{tabular}{lcc}
\toprule
\textbf{Ability Category} & \textbf{Fleiss’ Kappa (FK)} & \textbf{Krippendorff/Kendall’s Tau} \\
\midrule
Intellectual/Learning & 0.396 & 0.535 \\
Reduced Mobility      & 0.175 & 0.246 \\
Neurodiversity        & 0.327 & 0.471 \\
Deaf                  & 0.309 & 0.444 \\
Autism                & 0.180 & 0.463 \\
Blind                 & 0.303 & 0.376 \\
Mental Health         & 0.292 & 0.458 \\
\bottomrule
\end{tabular}
\Description{Table 3. Fully described in the text.}
\end{table*}
Although these values fall in the fair to moderate range, they do not indicate methodological failure. Rather, they reflect the subjectivity inherent in identifying nuanced ableism, which depends on context, lived experience, and normalization of bias \cite{odeaOutBoundsFactors2015,ebleSlangSociabilityIngroup1996,croomSlurs2011,galinskyReappropriationStigmatizingLabels2013}. The variation in agreement is therefore itself a finding, highlighting the diversity of perspectives within disability communities and the limits of deterministic labels. This finding is also consistent with previous work that shows that crowdsourced social computing tasks often exhibit low or variable inter-annotator agreement \cite{mullerDesigningGroundTruth2021,bowkerSortingThingsOut1999,gordonDisagreementDeconvolutionBringing2021}.

\subsubsection{\textbf{Perceived Presence of Ableism}}

Despite variability in sentence-level annotation, the majority of participants perceived the overall stories as ableist. In five out of seven surveys, more than 50\% of participants agreed that the overarching story contained either explicit or nuanced ableism. This finding affirms the success of our story design in prompting critical reflection on subtle ableist tropes embedded in plot, character development, and narrative framing.

\section{\textbf{Study 2 - Comparing Human and AI Ableism Annotations}}

The purpose of Study 2 is to compare human ableism annotations and AI ableism annotations. To achieve this, we created a unique “human baseline” annotation for each story from the crowdsourced annotations collected  in Study 1. Next, we used GPT-4o to create seven AI annotations using the same format and steps the humans used in study 1. Finally, we surveyed 108 members of the disability community trained in DEI/diversity studies to judge the quality and agreeableness of the two annotators under the guise that both are AI-generated. The following sections detail the methods and results of this study. 

\subsection{\textbf{Annotation Creation}}

\subsubsection{\textbf{Human Baseline Annotations}}

Rather than treating human annotations as a fixed ground truth \cite{mullerDesigningGroundTruth2021,sureshFrameworkUnderstandingSources2021}, we constructed a human baseline annotation that reflects what members of the disability community could reasonably consider ableist. We included any phrase that at least two participants independently identified as ableist for a similar reason, producing a composite annotation that preserves points of consensus while capturing the diversity of participant judgments. This approach creates a representative narrative rather than privileging any single annotator.

\textbf{Explanations and corrections:} For sentences identified as ableist, we used a two-step, human-led, and AI-assisted process to synthesize participant responses. GPT-4o produced an initial draft summary from small data sets (n = 4-10) for each explanation and correction. We chose this model following existing research showing that GPT can support consistent condensation of small text sets under close human oversight \cite{jangConsistencyAnalysisChatGPT2023,shakilEvaluatingTextSummaries2024,zhangBenchmarkingLargeLanguage2024,liBenchmarkingImprovingGeneratorValidator2023}. We generated the starter summary using the prompt \textit{“summarize the provided data into a single explanation. Use no outside data or information”} for the explanation and \textit{“summarize the provided data into a single sentence. Use no outside data or information”}, for the correction. 

It is important to note that these drafts were never used directly. The research team compared each AI summary line-by-line with the full set of original annotations, \textit{manually editing} the output to preserve participants’ exact language and reasoning and to remove any claims not grounded in the source data. To maximize reproducibility, all edits utilized the following procedure: 
(1) retain exact participant wording wherever possible; 
(2) add only words or phrases present in the original annotations; 
(3) remove any content not grounded in participant data; and 
(4) resolve disagreements by inclusion rather than omission. 

These rules ensured that editing was consistent and minimized researcher discretion, and the AI model’s role was limited to producing a first-pass compression that reduced researcher cognitive load. The \textit{Overall Story Analysis} part of the annotation followed the same procedure.

\textbf{Identifications:} We retained all phrases that participants independently highlighted for the same rationale in their explanations. These identifications were added directly from participant responses without modification.

Finally, while the complete dataset of human annotations is included in the supplemental information and in our online repository (\href{https://doi.org/10.5281/zenodo.18386877}{link}), the subset omitted from the synthesized human baseline is tagged to ensure transparency \cite{smith2026supplementary}. Future work can expand on this dataset to train culturally responsive bias indicators, while maintaining accountability to the voices of the participants who generated it.

\subsubsection{\textbf{Implications of AI-Assisted Synthesis on the Human Baseline}}

While this human-led and AI-assisted approach helped us efficiently manage qualitative synthesis, we recognize that the use of generative AI in constructing a “human” baseline is not without trade-offs. On one hand, having our small research team manually merge over 200 annotations into 7 different baselines would have introduced a substantial risk of researcher interpretation overshadowing the nuanced perspectives of disabled participants \cite{liangEmbracingFourTensions2021,pierreGettingOurselvesTogether2021,mullerDesigningGroundTruth2021}. AI-assistance helped mitigate that risk by reducing direct researcher authorship in early synthesis. At the same time, LLM-generated summaries can introduce overgeneralization or subtle shifts in meaning, emphasizing the necessity of human oversight \cite{huangSurveyHallucinationLarge2025,yangHallucinationDetectionLarge2025,shakilEvaluatingTextSummaries2024,sunPromptChainingStepwise2024}. To safeguard participant voice, our team followed a 4-part procedure to remove hallucinations and ensure accurate reflection of the underlying annotations.

This approach represents an attempt at both preserving participants’ intent and reducing researcher influence within the constraints of real-world qualitative dataset development. In future work and with increased resources, we hope this pipeline evolves to instead compensate disabled community members to guide or directly perform the synthesis, aligning future iterations more fully with disability-centered research values.

\subsubsection{\textbf{AI Annotations}}

To create AI annotations, we experimented with various GPT-4o prompting strategies. Although we initially trained GPT-4o using curated anti-ableism datasets \cite{departmentofglobalcommunicationsDisabilityInclusiveCommunicationsGuidelines2022,americanpsychologicalassociationInclusiveLanguageGuide20233,americanpsychologicalassociationwashingtondistrictofcolumbiaPublicationManualAmerican2020,nationalcenterondisabilityandjournalismDisabilityLanguageStyle2021,steenGlossaryAbleistWords2024}, the resulting annotations were often inconsistent and overly critical. We found that a chained prompting approach \cite{wuAIChainsTransparent2022,sunPromptChainingStepwise2024} produced the most coherent results.

The final AI annotation protocol included the following two prompts:

\begin{enumerate}
    \item \textit{“Given the following passage, go through each sentence and highlight every single phrase or concept that is explicitly or subtly ableist. Provide a detailed explanation of why each is ableist and suggest a more inclusive correction.”}
    \item \textit{“What themes or elements in the overall story might be considered ableist? Briefly explain why, and provide a more inclusive way to frame the story.”}
\end{enumerate}

\textbf{On the difference between the AI and Human annotation instructions:} The purpose of this study is to investigate how AI naturally annotates ableist language rather than investigating how AI would simulate a person with a disability annotating disabled language. Due to this distinction, the instructions given for annotation differed between the AI and human case but the steps (identification, explanation, and correction) remained the same. 

In study 1, human participants received extra instructions such as: "Use an objective and educational tone like a spell checker would. Refrain from speaking from the 'I' perspective and giving personal examples/anecdotes.", and "Write only the exact words or phrases that appear in the sentence and separate phrases with commas." These instructions were necessary to ensure we had user clean data we could directly combine to create a central human baseline annotation. After the trial and error process we used to create the AI prompt, our resulting prompts yielded annotations that naturally and consistently followed the same instructions we gave to the human annotators. The resulting annotation sets provided the data necessary to compare the difference between organic AI ableism annotations and those of members of the disability community. 

Prompting the LLM with the same instructions as human participants or asking it to adopt a disability-centered perspective would likely yield a different style of annotation. Exploring how such perspective-taking instructions shape AI outputs represents an exciting direction for future work.

\subsection{\textbf{Participants}}

We recruited 110 participants who self-identified with the disability community reflected in their assigned survey. Each participant had at least a bachelor’s degree and had prior experience with diversity, equity, and inclusion (DEI) or disability-related training or coursework.

Participants completed one of seven survey tracks based on disability identification: Vision Impairment, Hearing Impairment, Mental Health Condition, Intellectual or Learning Disability, Neurological Disability, Autism, and Physical Disability / Reduced Mobility. Each of the seven surveys had 15-16 responses. The most common age range among participants was 26–35 years, followed by 36–45 years. The majority identified as female (60), followed by male (41) and non-binary or third gender (5).

A large proportion of participants (73.6\%) identified with more than one disability. The most frequently reported disabilities were mental health conditions (76), physical disabilities or reduced mobility (46), and autism (26). 

Participants received \$6 for completing the 20-minute survey. Those who spent additional time were eligible to receive up to a \$2 bonus for their effort.

\begin{table}[h]
\centering
\caption{Participant Demographics (N = 106)}
\label{tab:participants2}
\begin{tabular}{lp{4.5cm}}
\toprule
\textbf{Category} & \textbf{Participant Data} \\
\midrule
Total Participants & 106 \\
\addlinespace[0.5em]
Age Distribution &
18--25: 13 \newline
26--35: 35 \newline
36--45: 26 \newline
46--55: 19 \newline
56--65: 9 \newline
66+: 4 \\
\addlinespace[0.5em]
Gender Distribution &
Female: $\sim$57\% (60) \newline
Male: $\sim$37\% (41) \newline
Non-binary/third gender: 5\% (5) \newline
Other: 0\% \newline
Prefer not to say: 0\% \\
\addlinespace[0.5em]
Disability Distribution &
Mental Health Condition: 76 \newline
Physical Disability/Reduced Mobility: 46 \newline
Autism: 26 \newline
Vision Impairment: 31 \newline
Neurological Disability: 33 \newline
Intellectual or Learning Disability: 20 \newline
Hearing Impairment: 23 \\
\addlinespace[0.5em]
Percentage who Identify\\ with Multiple Disabilities & 73.58\% \\
\bottomrule
\end{tabular}
\end{table}

\subsection{\textbf{Study Design}}

We used Qualtrics to create a survey where users reviewed the human and AI annotations and ranked which one they liked the best. First, participants signed an informed consent form and reviewed the same definitions of ableism and nuanced ableism used in Study 1. We told the participants to keep those definitions in mind while they read the story probe relating to their disability community. Next, they were randomly introduced to one of the annotators (human or AI), and read a preview of the story with the ableism phrases from the annotator in bold.

The rest of the survey progressed as follows:

\begin{enumerate}
    \item Participants viewed the original paragraph with all ableist phrases identified by the annotator in bold.
    \item For each sentence in the story with reported ableism, participants used a 5-point Likert scale to rank:
    \begin{enumerate}
        \item Whether the phrases highlighted by the annotator contributed to ableism in the sentence.
        \item How strongly they agreed with the annotator’s explanation.
        \item How strongly they agreed with the annotator’s suggested correction.
    \end{enumerate}
    \item Participants re-read the paragraph with all of the annotator’s revisions and rated how effectively the changes reduced ableism using a Likert scale.
    \item Participants reviewed the annotator’s passage-level analysis of ableism in the story and used another 5-point Likert scale to rank:
    \begin{enumerate}
        \item The validity of the identified themes.
        \item The strength of the explanation.
        \item The effectiveness of the alternative framing.
    \end{enumerate}
    \item Finally, participants answered questions about what they learned from the annotation and their perception of the annotator’s strengths, weaknesses, and areas for improvement.
\end{enumerate}
After repeating this process for a second annotator, participants indicated which annotator they preferred overall. The order of the two annotators was randomized. The participants were then informed that one of the annotators was based on crowd-sourced feedback from real disabled people. The final survey question inquires whether this information changed participants’ perceptions.

\section{\textbf{Results - Study 2}}

\subsection{\textbf{Quantitative Analysis: Do Participants Prefer the Human or AI Annotations?}}
\begin{figure*}
    \centering
    \includegraphics[width=0.8\linewidth]{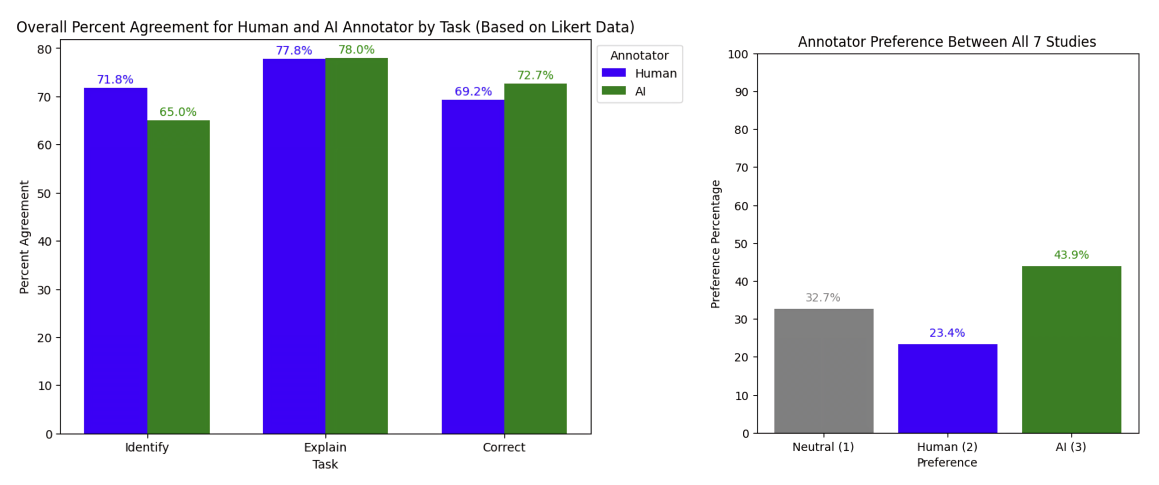}
    \caption{The graph on the left describes the percentage of times participants agreed or somewhat agreed with the human or AI identification, explanation, or correction of ableist language across all surveys. The graph on the right shows the percentage of people who preferred the AI ableism annotations, human ableism annotations, or those who had no preference.}
    \label{fig:placeholder}
    \Description{Figure 2. Fully described in the caption and results.}
\end{figure*}
Participants showed a high degree of agreement with both annotators when individually scoring their agreement with the identifications, explanations, and corrections by sentence. For both annotators, sentence-level agreement averaged across all three tasks was 72.3\%. A two-sample Z-test comparing average Likert scores for the two annotators confirmed that agreement levels for all three annotation steps were effectively equivalent: identification (Z = 1.47, \textit{p} = 0.14), explanation (Z = -0.05, \textit{p} = 0.96), and correction (Z = -0.75, \textit{p} = 0.45). Comparing the three annotation types separately, the human annotator was rated slightly higher in identification tasks in 6 out of 7 studies, while the AI was favored in 6 out of 7 studies for correction tasks. Preferences for explanations were evenly split.

However, when asked to choose their preferred annotator, participants significantly favored the AI annotator. The AI was preferred by 43.9$\%$ of respondents, compared to 23.4$\%$ for the human annotator, with 32.7$\%$ expressing no preference. A chi-square test (x² = 6.80, \textit{p} = 0.0333) and a two-sample Z-test for proportions (\textit{p} = 0.0011) confirmed this preference was statistically significant. This preference gap suggests that while participants judged both annotators similarly on a sentence-by-sentence basis, they may have perceived the AI’s annotations as more coherent, better formatted, or easier to interpret in aggregate.

These findings (summarized with Fig. 2) highlight that annotation “quality” is judged on more than correctness. Factors such as clarity, tone, and cognitive load appear to shape overall preferences, suggesting that even small stylistic choices can influence perceptions of credibility and usefulness. Additionally, preliminary subgroup analysis showed lower agreement in autism, blindness, and Deafness stories, especially for identification tasks. These cases may involve greater ambiguity or contested representations, underscoring the need for culturally grounded annotation practices. Overall, while human and AI annotators performed similarly on accuracy, participants favored the AI for its style and structure. Future refinement should build on these preferences while ensuring alignment with diverse disability perspectives.

\subsection{\textbf{Qualitative Style Analysis: What qualities of AI and human ableism annotations make them agreeable or disagreeable to the participants?}}

\subsubsection{\textit{\textbf{Procedure}}}

To come closer to answering \textit{RQ2: "What qualities of AI and human ableism annotations make them agreeable or disagreeable to the participants?}", we must first understand the style and tone differences between the human and AI annotations. We do this through a thematic coding with team that consisted of the first author and two independent contractors, both of whom were trained in cognitive science, psychology, and user studies. We reviewed the human and AI annotations for each  survey and used an inductive technique to converge on the central traits of the annotations. Each reader independently reviewed the responses to identify stylistic themes, patterns, and notable deviations. We held regular meetings to discuss the stylistic themes we found and supported them with text from each annotator. 

\subsubsection{\textit{\textbf{Human Annotation Style}}}

The human annotation adopts a justice-oriented style grounded in disability advocacy. The tone is assertive and unambiguous, using terms such as “inspiration porn,” “infantilization,” and “objectification” to label harmful patterns and convey moral urgency. A defining feature is its emphasis on social and political critique in explanations. For example, when analyzing “Emma, a talented software engineer with autism,” the annotation identifies the person-first phrasing as misaligned with community preferences. This depth of interpretation recurs throughout the annotations and favors decisive rewrites that shift narrative control back to the disabled character.

Corrections actively reframe story fragments to remove stereotypes while introducing affirming language. For example in a story about an autistic software developer, the sentence “Her colleagues, impressed by her ingenuity and precision, gathered to learn from her approach” was revised to “Her colleagues were happy that she was able to fix the problem, and she offered to teach them how to do it so they could learn themselves,” reframing the scene as one of mutual learning rather than putting someone on a pedestal because of their abilities. 

The passage-level analyses at the end of each human annotation set used essay-like prose to explain observations about ableist themes, often drawing on real-world implications and values rooted in lived experience. These sections summarize the overarching forms of ableism present in the text, such as excessive praise for basic tasks or the portrayal of disability as a “superpower”, and offer philosophical guidance on how to represent disabled individuals with dignity. 

\subsubsection{\textbf{AI Annotation Style}}

The AI annotations take a methodical, neutral approach, favoring speculative phrasing such as “may reinforce stereotypes” or “could be interpreted as” over firm judgments. The general tendency is to to evaluate language through the lens of potential misinterpretation rather than direct harm. It's corrections are typically conservative, softening or removing modifiers like “meticulously” or “quiet yet profound” while preserving sentence structure and authorial intent. The corrections are also often explained with short rationales. For example, it changes “Emma’s exceptional focus and pattern-recognition skills” to “Emma’s keen focus and problem-solving skills,” clarifying that the wording highlights professional strengths without attributing them stereotypical autistic traits.

Finally, the AI’s passage-level analyses are formatted in a structured, bullet-pointed style, categorizing each ableist trope with a “Why it’s ableist” explanation followed by an “Inclusive alternative.” For example, one section identifies “framing autism as the source of success” as problematic, then provides a clearly delineated alternative that emphasizes professional expertise over diagnostic traits. This structure gives the thematic section a reference-guide feel that is concise, modular, and easily skimmable for educational or editorial use.

\subsubsection{\textbf{Comparison and Implications}}

Both annotations address ableist bias through different strategies. The human annotator emphasizes cultural critique and empowerment, linking language to systemic issues and rewriting narratives to affirm disabled identity. The AI annotator, by contrast, adopts a neutral style: it avoids strong judgments, applies inclusive language principles, and favors edits that promote clarity and professionalism.

These orientations suggest different applications: the human style may resonate in community-led or DEI contexts, whereas the AI style may be better suited for editorial tools or scalable automation. These results combined with the direct participant feedback analyzed in the next section elaborates on these differences and what the disability community might prefer. 

\subsection{\textbf{Qualitative Analysis: Participant Feedback on Ableism Annotations}}

\subsubsection{\textbf{Procedure:}}

To analyze participants' open-ended responses, another thematic coding process was conducted with the same three researchers from the qualitative style analysis in the previous section. We used a deductive technique, looking specifically for strengths, weaknesses, and areas to improve within the data. Again, we held regular meetings to converge on a central coding of all of the annotations we read. In addition, opinions from the research team were cross-referenced with k-means clustering to extract other themes that we might have missed.

\subsubsection{\textbf{General Findings:}}

Participants provided rich and varied feedback on both the AI-generated and human-generated annotations of ableist language, revealing several trends with implications for the system’s design, interpretability, and future applications. Overall, the task was seen as \textbf{educational} and \textbf{thought-provoking}. Many participants shared that the annotations helped them recognize subtle or normalized ableist language, especially in stories they initially didn’t perceive as problematic. As one user noted, “I didn’t think this story was ableist at all when I began to read it, but after looking over the AI notes, I could come to agree with some of the things it was saying" (p7, Autism community). Others described the exercise as personally \textbf{meaningful:} “Even though I have been living with a disability for a few years now, I still have much of the negative concepts of disabilities in my head… I could see more easily why the wordings would be problematic and how best to approach amending them” (p79, Neurodiversity Community).

The AI annotator was often preferred for its consistency and clarity, particularly in the alignment between explanations and corrections. However, participants also expressed \textbf{concerns about over-editing}, especially in fiction. Many emphasized that fictional stories, where ableism may be depicted intentionally to drive the narrative, are not always appropriate test cases. As one respondent cautioned, “Simplifying those depictions changes the overall meaning” (p56, Mental Health Community). These insights highlight the need for genre-aware annotation approaches and thoughtful framing of the task’s educational goals.

\begin{figure*}[h!]
    \centering
    \includegraphics[width=0.8\linewidth]{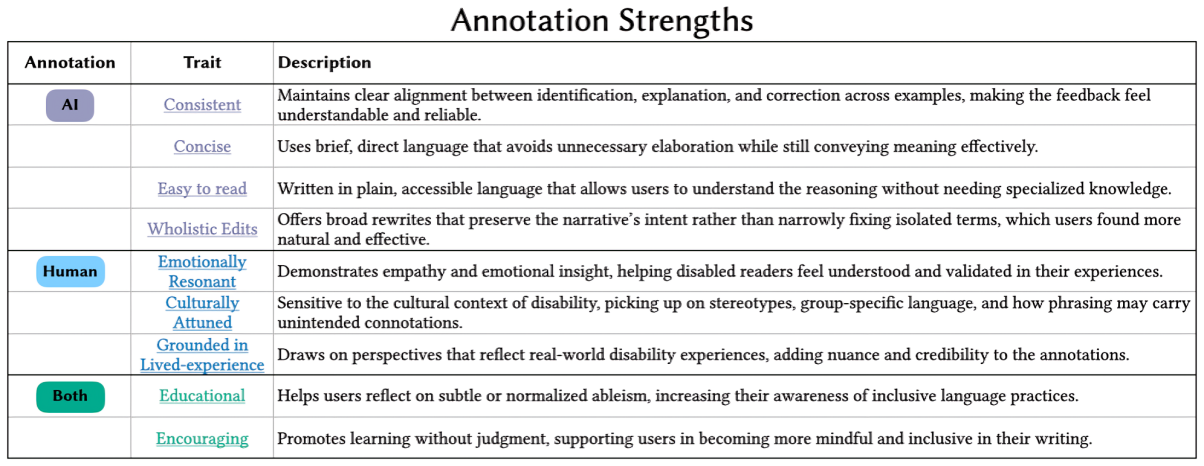}
    \caption{This table contains keywords and descriptions for the strengths of the human and AI annotators.}
    \label{fig:strength&weakness}
    \Description{Figure 3. This table contains a condensed version of the takeaways from section 5.3.3 Strengths.}
\end{figure*}

\subsubsection{\textbf{Strengths}}

\begin{itemize}
    \item \textbf{\textit{AI Annotations:}} Participants found both annotators valuable in helping them become more aware of ableist language. The AI annotator was praised for producing \textbf{concise, easy-to-read feedback}. One user remarked, “I liked that the explanations were in plain language and easy to understand” and appreciated that “the \textbf{corrections in general were as well-written as the original text}” (p12, Autism Community). The AI’s broader, more holistic edits were particularly well received: “This AI used much more broad strokes in the changes and \textbf{provided }\textbf{much more nuanced corrections} without sacrificing the initial intent” (p9, Autism Community). Others noted that it did a good job \textbf{defining unfamiliar terms} \textbf{and \textbf{identifying obvious forms of harm} without sounding judgmental} (p1, Autism Community; p17, Blind Community). A few participants commended its \textbf{consistency across examples} and its ability to \textbf{gently suggest better alternatives} (p8, Autism Community; p106, Reduced Mobility Community).
    \item \textit{\textbf{Human Annotations:}} For the human annotator, the strengths lay more in the \textbf{interpretive depth} of explanations, even if those strengths were diminished by inconsistent formatting or alignment with corrections. One participant shared, “Besides reducing the ableist language… it \textbf{taught me a lot} and is making me more aware of what to look for” (p11, Autism Community; p38, Deaf Community). Participants often highlighted the \textbf{emotional resonance} of the feedback, calling it “\textbf{thorough}” (p3, Autism Community), “\textbf{reflective}” (p44, Deaf Community), and \textbf{grounded in lived experience} (p1, Autism Community). Some praised the human annotator for modeling inclusive rephrasings in ways that \textbf{affirmed disabled identities}, not just avoided harm (p6, Autism Community; p40, Deaf Community). Others noted that the human feedback better \textbf{captured subtle cultural cues and stereotypes} that automated systems might overlook (p13, Autism Community; p21, Blind Community).
\end{itemize}

\begin{figure*}[h!]
    \centering
    \includegraphics[width=0.8\linewidth]{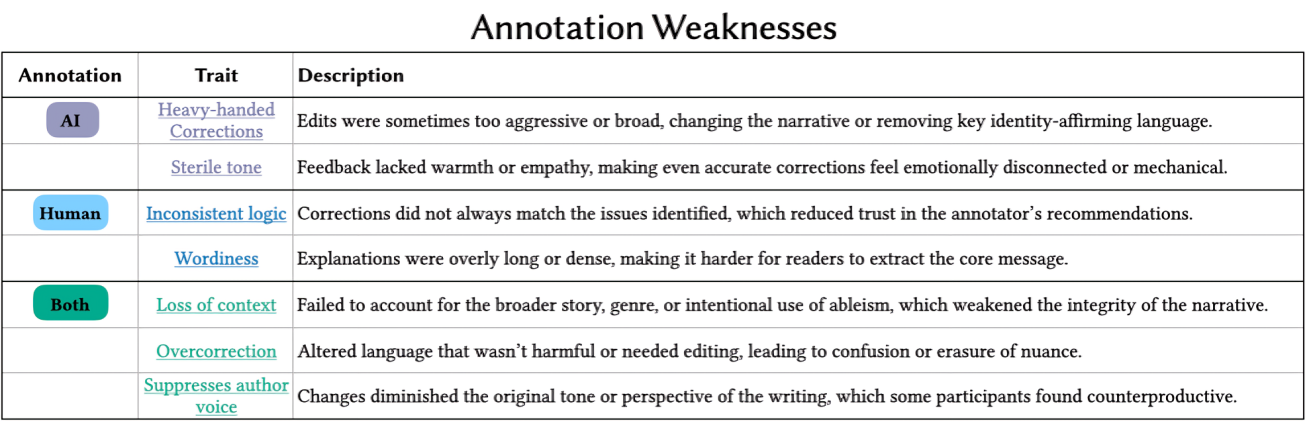}
    \caption{This table contains keywords and descriptions for the weaknesses of the human and AI annotators.}
    \label{fig:strength&weakness}
    \Description{Figure 4. This table contains a condensed version of the takeaways from section 5.3.4 Weaknesses.}
\end{figure*}

\subsubsection{\textbf{Weaknesses}}

\begin{itemize}
    \item \textbf{\textit{AI Annotations}}: Despite its strengths, the AI annotator occasionally \textbf{overcorrected} in ways that diminished narrative integrity or \textbf{erased important aspects of disabled identity}. For example, one user wrote that “removing mention of Deafness… diminished the story’s nuanced portrayal of the character’s experience” (p38, Deaf Community). Others felt that \textbf{line-by-line corrections} missed the “big picture” or stripped the story of \textbf{complexity and nuance} (p2, Autism Community; p26, Blind Community). Several participants described the AI’s tone as sometimes \textbf{sterile or lacking empathy}, which led to corrections that felt technically accurate but emotionally disconnected (p12, Autism Community). A few also mentioned that some of the \textbf{edits disrupted the} \textbf{clarity }or flow of the writing, making the narrative feel mechanical or over-sanitized (p9, Autism Community; p32, Deaf Community).
    \item \textbf{\textit{Human Annotations}}: The human annotator, by contrast, was criticized for \textbf{inconsistent logic} between identified ableism and the suggested corrections. Several participants noted that the revisions did little to reduce bias: “The corrections went against their explanations and did not change the paragraph much to make it less ableist” (p26, Blind Community). Others commented on the \textbf{dense, wordy format}: “It is quite wordy in its explanations… improvement could come from a more engaging format” (p31, Blind Community). Some participants also flagged issues with \textbf{unclear phrasing} and \textbf{grammatical awkwardness}, which made the corrections harder to follow or implement (p17, Autism Community; p50, Mental Health Community). While the human feedback was often more emotionally resonant, these critiques suggest a \textbf{need for \textbf{stronger alignment} between explanation and correction}, as well as more attention to structure and readability.
\end{itemize}

\begin{figure*}[h!]
    \centering
    \includegraphics[width=0.8\linewidth]{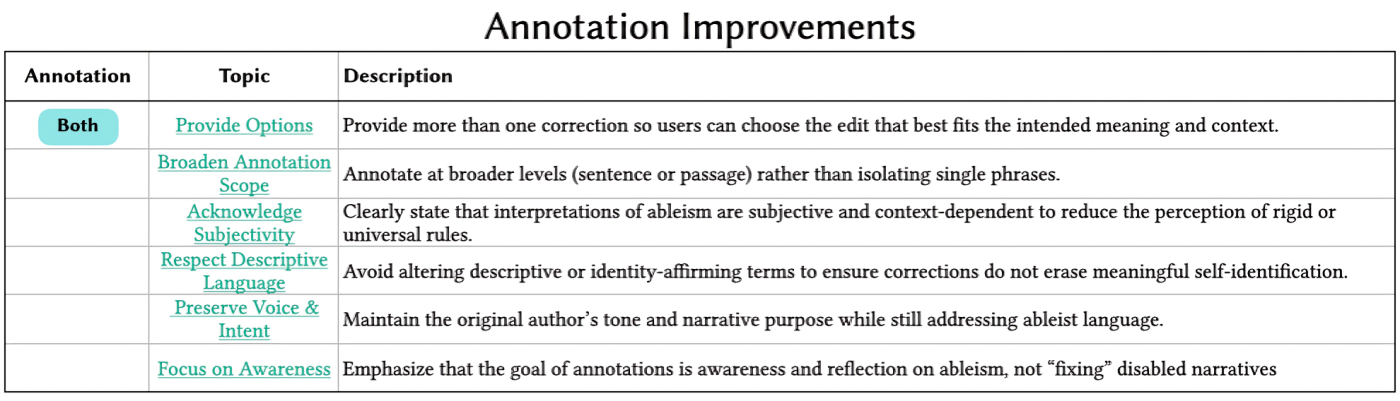}
    \caption{This table contains keywords and descriptions for what participants believed should improve about the ableism annotation system used in this study.}
    \label{fig:strength&weakness}
    \Description{Figure 5. This table contains a condensed version of the takeaways from section 5.3.5 Improvements.}
\end{figure*}

\subsubsection{\textbf{Improvements}}
 \mbox{}\\[0.2em]

Participants proposed several changes that could make the annotation process more useful and respectful. These included \textbf{providing multiple correction options}, \textbf{annotating at the sentence or passage level} rather than flagging individual phrases, and \textbf{explicitly stating the subjective nature of ableist interpretations}. A recurring suggestion was to \textbf{avoid altering descriptive or identity-affirming language}. As one user put it, “Quit taking away from what was meant to be descriptive (p36, Deaf Community).” Others emphasized the importance of \textbf{preserving voice and intent}: “More work needs to be done to keep the integrity of what’s trying to be said intact. (p2, Autism Community)”

Several participants called for \textbf{deeper inclusion of disabled voices}, particularly when annotating language related to neurodiversity: “Unless you’ve lived in an autistic mind, you really shouldn’t have much of an opinion" (p12, Autism Community). There was also a call for better framing of the task’s goals, as some dismissed the project as unnecessary: “People with disabilities are a lot more resilient than you’re giving us credit for" (p60, Mental Health Community). These comments suggest that future iterations must \textbf{clearly communicate the purpose of the annotations}—not to “fix” disabled narratives, but to foster awareness of ableism in everyday language (p49, Mental Health Community).

Taken together, the feedback affirms the tool’s potential as a socially responsive educational resource. However, its continued success will depend on greater contextual sensitivity, participatory design, and framing that invites reflection without erasure. These priorities will be essential for building a system that supports inclusion while respecting diverse forms of expression.

\subsection{What value do ableism annotations hold for disabled communities and beyond?}

Participant feedback in this study reveals a wide range of educational, reflective, and emotional responses to the annotation process, highlighting the potential value of ableism annotations both within and outside of disabled communities. While many participants appreciated the opportunity to better understand microaggressions and refine their communication, others voiced important concerns about erasure, simplification, and outsider authority. When thoughtfully framed, annotation tools can foster awareness and inclusion, but they must be designed with caution, humility, and community accountability.

\subsubsection{\textbf{Societal Benefits: Educational Value and Awareness Building}}

Across disability identities, participants said the annotations helped them recognize subtle ableism they had overlooked and gave language to long-felt concerns. One reflected, \textit{“It defined language that always bothered me, but I couldn't explain why”} (p1, Autism Community). Others described shifts in self-awareness: \textit{“I learned that I can be ableist without even realizing it”} (p26, Blind Community), and \textit{“Now, I am aware that I may be patronizing them and will be more careful with my wording”} (p20, Blind Community).

Many emphasized the broader value of the tool. As one participant noted, \textit{“If I hadn't already been familiar with this subject matter, I do think the AI provided clear, persuasive reasons why this subtle language was ableist… [It] could change the mind of people who weren't previously very aware of the topic”} (p14, Autism Community). Another wrote, \textit{“I don't see any more to improve except for this to spread to every workplace environment”} (p10, Autism Community). Participants also highlighted how the task revealed diversity within disability communities, with one sharing, \textit{“It taught me to take in how others might feel about their life with their difficulty”} (p20, Blind Community).

\subsubsection{\textbf{Risks of Erasure and Overreach}}

While many participants valued the task, others cautioned against misusing ableism annotations. A recurring theme was the need to ground decisions in lived experience. As one participant warned, \textit{“Unless you've lived in an autistic mind, you really shouldn't have much of an opinion… neurotypical individuals should not be in charge of writing things and creating rules on behalf of anyone who is autistic”} (p12, Autism Community).

Some criticized efforts to enforce a single “acceptable” language, arguing this could flatten complex identities: \textit{“Stop trying to force people to use one set of ‘acceptable’ language… In a way it targets those with disabilities as victims that need saving”} (p37, Deaf Community). Participants in the mental health community raised additional concerns about over-sanitization. \textit{“It's an easy way out—changing the language rather than confronting what people are actually going through”} (p48, Mental Health) and \textit{“Diluting it is ableist because it reduces the struggle of the people going through it”} (p59, Mental Health Community). Risks of erasure and overreach rise when use of bias annotation tools (and related toxicity detection tools \cite{phutaneColdCalculatedCondescending2025}) are not aligned with the proper use case. Potential solutions to ensure this alignment, along with the implications and impacts of this work are explored later in the Discussion and Future Work sections. 

\section{Discussion}

This work contributes empirical insight into how AI systems engage with ableist language when evaluated directly by disabled communities. As disability studies and Disability Justice scholars emphasize, language is inseparable from cultural power and lived identity \cite{campbellContoursAbleismProduction2009,berneTenPrinciplesDisability2018}. While prior HCI and NLP research has demonstrated that AI systems both reproduce ableist assumptions and struggle to identify implicit harms \cite{narayananvenkitAutomatedAbleismExploration2023,phutaneColdCalculatedCondescending2025}, little is known about how disabled people perceive AI-generated explanations and corrections, or how those preferences compare to community-grounded human annotations.

Our findings suggest that AI can serve as a helpful collaborator in bias education—particularly through consistent formatting, accessible tone, and actionable revisions—yet still falls short in preserving narrative nuance and affirming disabled identity. By studying annotation performance in narrative contexts, which play a key role in how ableism is communicated and understood \cite{keller_microaggressive_2010}, this work extends prior efforts focused mainly on short-form or decontextualized text. The following sections reflect on the implications of these results across our three research questions and offer design guidance for AI systems aimed at supporting inclusive writing and cultural competence.

\subsection{\textbf{RQ1: Do humans prefer AI annotators or human annotators for identifying, explaining, and correcting ableist language?}}

Participants showed no significant difference in agreement with AI versus human annotations across identification, explanation, and correction (average agreement 72.3\%). Both were perceived as equally accurate. When asked to choose a preferred annotator overall, however, participants significantly favored the AI ($\chi^2$ = 6.80, \textit{p} = 0.0333), citing its consistency, clarity, and accessible formatting. Still, fewer than half selected the AI outright, suggesting broad comparability in perceived quality.

These results carry important implications for the deployment of annotation methods in research and practice. The comparable accuracy between AI and human annotators demonstrates that integrating AI-based tools into the annotation process can maintain methodological rigor without a trade-off in performance. Additionally, the significant preference for the AI’s narrative style and formatting suggests that features beyond numerical accuracy, like clarity, consistency, and accessibility, can meaningfully influence user perceptions. Future annotation systems might benefit from a hybrid approach, leveraging AI’s strengths in standardization and presentation while incorporating nuanced human oversight for context and sensitivity. Ultimately, these insights contribute to a more nuanced understanding of how technology can augment traditional human-centered practices in detecting and addressing ableist language, and they highlight avenues for further research into optimizing both the technical and narrative elements of automated annotation systems.

\subsection{\textbf{RQ2: What qualities of AI and human ableism annotations make them agreeable or disagreeable to the participants?}}

Participants found both annotators educational but valued different strengths. The AI was praised for clarity, neutrality, and formatting that made corrections easy to apply, though some criticized it as emotionally detached or prone to over-sanitizing disability representation. The human annotator was appreciated for cultural grounding, advocacy language, and narrative-level critique, yet was faulted for inconsistent logic and dense phrasing.

These responses suggest that agreeableness is not just a function of correctness, but of how well the annotation aligns with a reader’s values, expectations, and sense of narrative integrity. Where the AI excels in editorial alignment, the human annotator adds cultural context and affective resonance. Disagreeable annotations, by contrast, often missed the mark in either clarity or cultural sensitivity, or failed to maintain a coherent relationship between explanation and correction.

\subsection{\textbf{RQ3: How can AI annotators improve at identifying, explaining, and/or correcting ableist language?}}

Participant feedback highlights four key directions for improving AI annotators. First, AI systems must strengthen contextual sensitivity. Users noted that the current model sometimes removes identity-affirming or narrative-relevant details, particularly in fictional settings, which can inadvertently erase lived experience rather than address bias. Second, improvements are needed in alignment between explanations and corrections. When the two were inconsistently connected, participants questioned the legitimacy and usefulness of the edits. Future systems should ensure each revision clearly resolves the identified issue. Third, participants emphasized preserving voice and intent. Corrections that felt overly literal, mechanical, or sanitized disrupted narrative flow. Providing multiple correction options and further integrating sentence-level and passage level edits (rather than presenting them in isolation) would help maintain storytelling integrity. Finally, users called for greater inclusion of disabled perspectives in how decisions are made and framed. Clearer communication that annotations aim to foster awareness—not prescribe a single “correct” vocabulary—will be essential to avoid overreach and maintain trust. Overall, these findings indicate that the most impactful improvements will come from tools that balance AI’s strengths in clarity and structure with flexible, culturally grounded decision-making and deeper community involvement in design.

\subsection{\textbf{Guidelines for developers of LLMs and writing tools that deal with culturally sensitive language}}

\begin{figure*}[h!]
    \centering
    \includegraphics[width=0.4\linewidth]{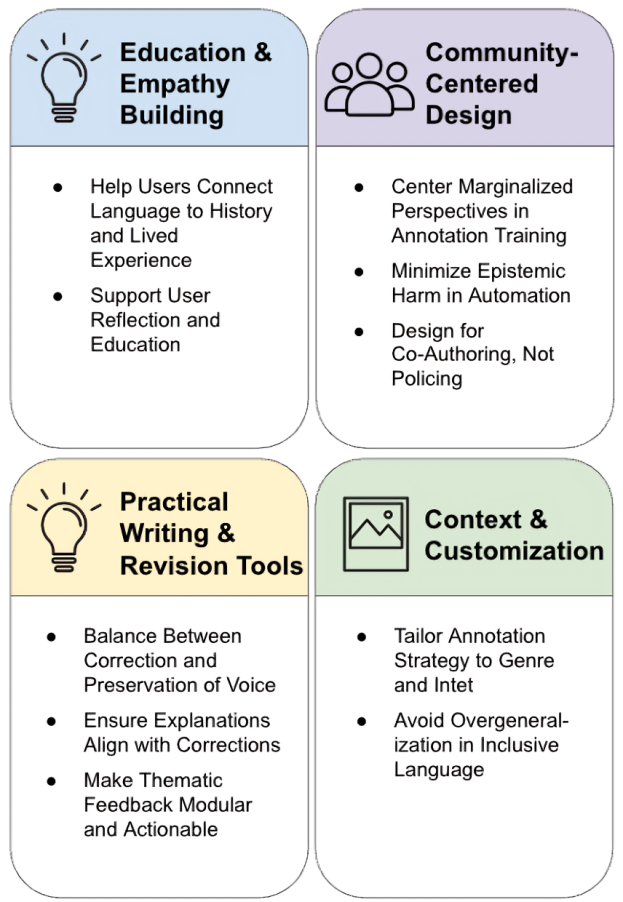}
    \caption{Summary of Guidelines for developers of LLMs and writing tools that deal with culturally sensitive language}
    \label{fig:guidelines}
    \Description{Figure 6. This figure is fully described in section 6.4.}
\end{figure*}
Drawing from participants’ reflections on the strengths, weaknesses, and suggested improvements of both human and AI annotators, we present a set of foundational guidelines for future research. These recommendations are organized into four thematic pillars, reflecting core priorities for developers building annotation systems that address culturally sensitive content. We hope that integrating some of these guidelines into the prompt and design of AI bias detection and correction tools will create tools that are more aligned with the needed of underrepresented communities while educating intended users.

\textbf{ 1) Education and Empathy Building}
    \begin{itemize}
        \item \textbf{ Help users connect language to history and lived experience}
            \begin{itemize} 
                \item   Many participants described the annotation task as educational, noting that it helped them reconsider language they previously viewed as neutral. To build empathy, AI systems should ground their feedback in historical and cultural context by explaining why certain terms carry harmful implications and how they relate to the lived experiences of marginalized communities \cite{garland-thomsonFeministDisabilityStudies2005,hassanUnpackingInterdependentSystems2021,berneTenPrinciplesDisability2018}. This encourages users to move beyond rule-following and engage in deeper reflection.
            \end{itemize}
    \end{itemize}
    \begin{itemize}
        \item \textbf{Support user reflection and education}
        \begin{itemize} 
            \item Rather than offering directives, annotations should act as prompts for self-awareness. Several participants suggested that corrections be framed as suggestions, not mandates, to invite personal interpretation and promote inclusive communication as an ongoing learning process. Annotations that foster curiosity rather than compliance are more likely to produce long-term behavioral change \cite{bucincaTrustThinkCognitive2021}.
        \end{itemize}
    \end{itemize}
\textbf{2) Community-Centered Design}
\begin{itemize}
    \item \textbf{Center marginalized perspectives in annotation training}
    \begin{itemize}
    \item Participants emphasized that annotations should reflect the values, insights, and language preferences of the communities they aim to serve. Feedback such as “unless you’ve lived in an autistic mind…” touch on the need to involve disabled people not just as annotators, but as collaborators in system design. Models must be informed by community-grounded data and validated through participatory processes \cite{gadirajuWouldntSayOffensive2023,pierreGettingOurselvesTogether2021}.
    \end{itemize}
    \item \textbf{Minimize epistemic harm in automation}
    \begin{itemize}
    \item Epistemic harm occurs when dominant perspectives override marginalized ones, or when tools frame bias as objective fact \cite{pierreGettingOurselvesTogether2021,sureshFrameworkUnderstandingSources2021}. Several participants expressed concern that AI annotations, if not carefully designed, might erase identity-affirming language or suggest overly normative corrections. Systems should clearly acknowledge the subjectivity of ableist interpretations and avoid treating all feedback as universally applicable.
\end{itemize}
\end{itemize}
\begin{itemize}
    \item \textbf{Design for co-authoring, not policing}
    \begin{itemize}
    \item Annotation systems should act as writing partners, not enforcers. Participants were more receptive to feedback when it was presented as collaborative and respectful, rather than corrective or judgmental. Co-authoring means offering alternatives, clarifying intentions, and preserving agency instead of stripping a story of its meaning.
\end{itemize}
\end{itemize}
\textbf{3) Practical Writing \& Revision Tools}
\begin{itemize}
    \item  \textbf{Balance between correction and preservation of voice}
    \begin{itemize}
        \item Participants flagged cases where the AI’s well-intentioned edits removed valuable narrative details or flattened identity expressions. Effective writing tools must distinguish between harmful bias and meaningful self-description. Where possible, annotations should preserve tone, character, and voice, especially in fiction or lived-experience writing.
    \end{itemize}
    \item \textbf{Ensure explanations align with corrections}
    \begin{itemize}
        \item Several respondents noted a mismatch between some explanations and revised sentences in the annotations. To maintain credibility and usefulness, each correction should be accompanied by a clear rationale that ties directly to the identified issue. Misaligned explanations and corrections undermine trust and reduce the educational value of the tool.
    \end{itemize}
    \item \textbf{Make thematic feedback modular and actionable}
    \begin{itemize}
        \item Participants valued feedback that was clear and easy to implement. They suggested formatting thematic feedback as modular blocks (e.g., bullet points or sections) so users can easily refer to specific ideas. This style makes the feedback more digestible and helps users prioritize which changes to make.
    \end{itemize}
\end{itemize}
\textbf{4) Context and Customization}
\begin{itemize}
    \item \textbf{Tailor annotation strategy to genre and intent}
    \begin{itemize}
        \item Participants cautioned against applying the same annotation logic across all types of writing. For example, fictional or satirical texts may include ableist tropes for narrative purposes, and editing them out can distort meaning. Systems should be able to adapt their tone, framing, and thresholds for correction depending on genre and authorial intent.
    \end{itemize}
    \item \textbf{Avoid overgeneralization in inclusive language}
    \begin{itemize}
        \item Several participants pointed out that terms commonly flagged as ableist may be identity-affirming within specific communities. Systems should not rely on static word lists or rigid rules. Instead, they should evaluate language in context, allowing for flexible interpretations that recognize intra-group variation and cultural nuance \cite{ebleSlangSociabilityIngroup1996,galinskyReappropriationStigmatizingLabels2013,nationalcenterondisabilityandjournalismDisabilityLanguageStyle2021}.
    \end{itemize}
\end{itemize}

These guidelines reflect a broader shift from algorithmic policing toward collaborative meaning-making. By centering user reflection, community values, and contextual awareness, AI systems can move beyond mere detection and become tools for inclusive, participatory storytelling.

\section{Limitations and Future Work}

\subsection{\textbf{Evaluating ableism bias in a larger range of literary scenarios: }}
Our study evaluated annotations in short fictional stories, carefully designed to include subtle ableist language. Participants noted, however, that ableism in fiction is framed differently than in real-world contexts. Tropes may be used deliberately for critique or realism, and editing them without regard to genre risks distorting meaning. As such, findings may not fully generalize to nonfiction, autobiographical, or professional writing. Future work should test annotation strategies across domains such as medical writing, journalism, academic prose, and lived experience narratives.

\subsection{\textbf{Expanding ableism datasets to include more disabled voices: }}
To capture in-group expertise, our participant pool included college-educated disabled people with prior training in ableism and bias. While deeply informed, this group was not representative of the broader disability community. Although there is consensus that marginalized voices should guide this work, it remains unclear who within those communities should be included or excluded when constructing “expert” systems \cite{liangEmbracingFourTensions2021,birhanePowerPeopleOpportunities2022,cooperSystematicReviewThematic2022}. Future work should test how annotation outcomes change with more diverse participation and fewer restrictions on who contributes.

\subsection{\textbf{Use crowdsourced ableism data to improve AI bias annotation models:}}

A key limitation of this study is that our human baseline involved editorial synthesis: we used AI-assisted drafts and then manually refined them to preserve participant voice. While this raises reproducibility concerns, we mitigated them by applying a documented set of editing rules (retain participant wording, add only from source data, remove unsupported claims, and resolve disagreements by inclusion) and by tagging all annotations omitted from the final baseline for transparency. In this sense, the baseline is reproducible as a rule-governed qualitative synthesis rather than an exact replication of each editorial decision.  

At the same time, this work contributes the first crowdsourced dataset of nuanced ableism annotations, but the scale is still limited. Larger datasets will be needed to reliably train bias-aware AI models that can generalize across contexts. While it is unfair to place the burden of education solely on marginalized groups \cite{chordiaSocialJusticeHCI2024,liangEmbracingFourTensions2021,pierreGettingOurselvesTogether2021}, community-generated data can both guide model training and reduce the expectation that individuals continually serve as one-on-one educators. Expanding this dataset and using them to train more bias education AI tools could empower allies to learn from authentic perspectives while easing the educational burden on those most affected.

\subsection{\textbf{Exploring the long term effects of educational bias annotations:}}
 A core motivation for this work is to develop tools that support everyday anti-bias practice. Future research should assess not only immediate learning, but also retention, behavioral change, and long-term effects on writing. It is also important that learning and behavioral changes be studied with both members of the target community and non-members who would use this tool. Participants also expressed a desire for tools that encourage reflection and offer choices rather than prescriptive corrections, and this could also improve participant learning in future studies. Positive shifts in language after using this tool use would signal the value of annotation tools for both members of marginalized communities and nonmembers seeking to build more inclusive habits.

\subsection{\textbf{Expanding this framework to other forms of bias:}}
Most importantly, this work lays the foundation for a broader, scalable framework to address biased language across many forms of marginalization. Ableism is just one axis of bias, and our long-term goal is to develop a generalizable annotation process that can be applied to language regarding race, gender, religion, economic status, and other identities. By collecting annotation data and feedback from multiple underrepresented communities, we aim to train models that are capable of recognizing and responding to nuanced bias in a socially and culturally aware manner. Ultimately, we envision a future where AI systems serve as collaborative partners in inclusive writing, grounded in community knowledge and capable of addressing intersectional forms of harm.

\section{Conclusion}

This paper introduces a framework for creating community-informed and educational bias annotators, showing that AI systems can identify, explain, and correct ableist language with accuracy while supporting learning. Through a two-part study, we contribute (1) a first-of-its-kind dataset of nuanced ableism annotations rooted in lived experience, (2) an empirical comparison of AI- and human-generated annotations that highlights the tradeoffs between consistency, clarity, and cultural depth, and (3) design guidelines for writing tools that promote inclusive communication while respecting narrative integrity and community values. AI annotations were strong in clarity and educational tone, while human annotations excelled delivering cultural insight, and advocacy-oriented framing. Ultimately, participants’ strong preference for the AI suggests that with improvements it could become a reputable bias educator, reducing the burden on marginalized communities to educate others while empowering allies to learn. We hope that future work can utilize our annotation framework and our ableism dataset to create culturally competent AI tools that can educate, support, and scale inclusive practices while remaining accountable to those most affected by bias.

\section{Appendix}
\appendix
\section{Nuanced Ableism Definition}

This study is focused on identifying nuanced ableism in
text. Nuanced ableism refers to subtle, often unconscious,
forms of discrimination or prejudice against people with
disabilities. Unlike overt ableism, which is explicit and direct,
nuanced ableism can be harder to recognize and can
manifest in various ways, such as:

\begin{enumerate}
    \item Infantilization: Treating adults with disabilities as if they are children, which includes using baby talk or assuming they are incapable of performing tasks or making decisions without guidance.
    \item Inspiration Porn: Depicting people with disabilities as inspirational solely or primarily because of their disability, often used to make non-disabled people feel grateful or inspired.
    \item Benevolent Ableism: Offering help or making accommodations without asking if it’s needed, assuming people with disabilities always need assistance.
    \item Assuming Incompetence: Assuming that a person with a disability is less capable or intelligent,often leading to speaking over them or making decisions on their behalf.
    \item Euphemisms and Avoidance: Using euphemisms like "differently-abled" or avoiding the topic of disability altogether, which can trivialize or stigmatize the experiences of disabled individuals.
    \item Condescension: Talking down to or patronizing people with disabilities, often through exaggerated praise or simplified language.
    \item Excessive Praise for Basic Tasks: Praising people with disabilities for completing ordinary tasks, which can be  condescending and imply low expectations.
    \item Assumptions about Quality of Life: Assuming that a person’s disability necessarily leads to a poor quality of life and expressing undue pity or sorrow.
    \item Assumptions about Interests and Abilities: Assuming people with disabilities cannot have interests, talents, or ambitions similar to those of non-disabled people.
    \item Treating Disability as Tragedy: Framing disability primarily as a tragedy or burden, which can perpetuate negative stereotypes and stigma.
\end{enumerate}

These forms of ableism can be subtle but are pervasive in
various aspects of society, often reinforcing stereotypes
and creating barriers for people with disabilities.

\section{Criterion for Ableist Stories}
The following evaluation criteria ensures that at least 3-4 examples of nuanced ableism are present in our selected stories. If a story failed two out of three of the items on this list, we modified the GPT prompt to include items 2 and 3 from this rubric directly. 

\begin{enumerate}
    \item Do at least half of the beta testers believe the story is ableist?
    \item Does the story contain at least 2 of the following common traits of nuanced ableist speech? [apa, UN guidelines, center for Disability Studies]
    \begin{enumerate}
        \item Portray them as having survived or battled their disability
        \item Mentioning a disability frequently
        \item Use euphemisms like differently abled, people of all abilities, people of determination, special needs/assistance
        \item Display their condition as a burden or problem that people w/o disabilities should solve or help with
        \item Refer to them by diagnosis
        \item Portray them as victim, or someone who suffers from, is afflicted with, or is stricken with a disability
        \item Referring to the person as stuck "inside" a body with the disability
    \end{enumerate}
    \item Does the story contain at least 1 phrase that is mentioned by reputable sources to be subtly ableist?
\end{enumerate}

\section{Example AI-Generated Story}
Prompt: \textit{"Write a 5 to 8 sentence short story about a blind person doing their job or completing a task."}

Story: “In the quiet of the early morning, Emily, who is blind, set about her usual routine of brewing her favorite cup of coffee. With practiced ease, she navigated her kitchen, relying on her heightened senses and memory. She reached for the coffee beans, her fingers lightly brushing over the Braille labels she had meticulously placed on each jar. After grinding the beans, she measured the right amount by touch and poured water into the kettle, listening carefully for the sound of boiling. As the aroma of freshly brewed coffee filled the room, Emily felt a sense of accomplishment. Her morning ritual, though simple, was a testament to her independence and adaptability, proving that with determination and skill, she could seamlessly manage daily tasks.” (GPT4-o)

The rest of the stories can be found in our supplemental information and in our online repository (\href{https://doi.org/10.5281/zenodo.18386877}{link}) \cite{smith2026supplementary}.

\begin{acks}
This work was conducted while author Kynnedy Simone Smith was an intern at Microsoft Research, mentored by Danielle Bragg. Special thanks to researcher Denae Ford for your invaluable help behind the scenes that allowed this work to be shared with the community. We also thank Claire Jong and Eliana Birman for assisting with the thematic coding for this study. Finally, we extend thanks to our study participants for sharing their time and their stories.
\end{acks}
\bibliographystyle{ACM-Reference-Format}
\bibliography{references}

\end{document}